\documentstyle[11pt,epsfig]{article}
\title{\bf Generalized Chaplygin gas as  geometrical dark energy}
\author{M. Heydari-Fard\thanks{email:
m.heydarifard@mail.sbu.ac.ir} \hspace{0.5mm} and H. R.
Sepangi\thanks{email: hr-sepangi@sbu.ac.ir}
\\ {\small Department of Physics, Shahid Beheshti University, Evin, Tehran 19839, Iran}}
\begin{document}
\maketitle 
\begin{abstract}
The generalized Chaplygin gas provides an interesting candidate
for the present accelerated expansion of the universe. We explore
a geometrical explanation for the generalized Chaplygin gas within
the context of brane world theories where matter fields are
confined to the brane by means of the action of a confining
potential. We obtain the modified Friedmann equations,
deceleration parameter and age of the universe in this scenario
and show that they are consistent with the present observational
data.\vspace{2mm}\noindent
\vspace{5mm}\\
\end{abstract}
\section{Introduction}

In recent times the generalized Chaplygin gas (gCg) model has been
proposed as an alternative to both the cosmological constant and
quintessence in explaining the accelerated expansion of our
universe. The Chaplygin gas model describes a transition from a
universe filled with dust-like matter to an accelerated expanding
stage. The generalized Chaplygin gas model, introduced in \cite{3}
and elaborated in \cite{4}, is described by a perfect fluid
obeying an exotic equation of state
\begin{eqnarray}
p=-\frac{B}{\rho^{\beta}},
\end{eqnarray}
where $B$ is a positive constant and $0<\beta\leq1$. The standard
Chaplygin gas corresponds to $\beta=1$. An attractive feature of
the model is that it can naturally explain both dark energy and
dark matter \cite{5}. The reason is that the Chaplygin gas behaves
as dust-like matter at the early stages of the evolution of the
universe and as a cosmological constant at late times. The
Chaplygin gas appears as an effective fluid associated with
$d$-branes \cite{6,7} and can also be derived from the Born-Infeld
action \cite{9}. An interesting range of models have been found to
be consistent with the SNe Ia data \cite{SNe}, CMB experiments
\cite{CMB} and other observational data \cite{other}. The
cosmological implications of the Chaplygin gas model have been
intensively investigated in recent literature
\cite{cosmology1,cosmology2}.

Over the past few years, models with extra dimensions have been
proposed in which the standard fields are confined to a
four-dimensional ($4D$) world viewed as a hypersurface (the brane)
embedded in a higher dimensional space-time (the bulk) through
which only gravity can propagate. The most popular model in the
context of brane world theory is that proposed by Randall and
Sundrum (RS). The so-called RSI model \cite{11} proposes a
mechanism to solve the hierarchy problem with two branes, while in
the RSII model \cite{12} a single brane with positive tension is
considered where $4D$ Newtonian gravity is recovered at low
energies even if the extra dimension is not compact. This
mechanism provides us with an alternative to the compactification
of extra dimensions.

The cosmological evolution of such a brane universe has been
extensively investigated and effects such as a quadratic density
term in the Friedmann equations have been found
\cite{13}-\cite{14}. This term arises from the imposition of the
Israel junction conditions which is a relationship between the
extrinsic curvature and energy-momentum tensor of the brane and
results from the singular behavior in the energy-momentum tensor.
There has been concerns expressed over applying such junction
conditions in that they may not be unique. Indeed, other forms of
junction conditions exist so that different conditions may lead to
different physical results \cite{16}. Furthermore, these
conditions cannot be used when more than one non-compact extra
dimension is involved. Against this background, an interesting
higher-dimensional model was introduced in \cite{17} where
particles are trapped on a 4-dimensional hypersurface by the
action of a confining potential ${\cal V}$. In \cite{18}, the
dynamics of test particles confined to a brane by the action of
such a potential at the classical and quantum levels were studied
and effects of small perturbations along the extra dimensions
investigated. Within the classical limits, test particles remain
stable under small perturbations and the effects of the extra
dimensions are not felt by them, making them undetectable in this
way. The quantum fluctuations of the brane cause the mass of a
test particle to become quantized and, interestingly, the
Yang-Mills fields appear as quantum effects. Also, in \cite{018} a
confining potential formalism was used to confine particles to an
arbitrary manifold in a higher dimensional Euclidean space. In
\cite{19}, a brane world model was studied in which the matter is
confined to the brane through the action of such a potential,
rendering the use of any junction condition unnecessary and
predicting a geometrical explanation for the accelerated expansion
of the universe. In a related work \cite{Gauss-Bonnet}, a brane
scenario was studied where the $m$-dimensional bulk is endowed
with a Gauss-Bonnet (GB) term and the localization of matter on
the brane is again realized by means of a confining potential. It
was shown that in the presence of the GB term, the universe
accelerates faster than brane models without the GB term. The
behavior of an anisotropic brane world with Bianchi type I and V
geometry was studied in \cite{anisotropic} along the same lines.

In this paper, we follow \cite{19} and consider an $m$-dimensional
bulk space without imposing the $Z_2$ symmetry. As mentioned
above, to localize matter on the brane, assumed to be thin, a
confining potential is used rather than a delta-function in the
energy-momentum tensor. The resulting Friedmann equations on the
brane are modified by an extra term that may be interpreted as the
generalized Chaplygin gas, providing a possible phenomenological
explanation for the accelerated expansion of the universe. It
should be emphasized that the model presented in this work is
different from that introduced in \cite{23,24} in that the latter
provides no account for the confinement of matter to the brane.
\section{The model}
In this section we present a brief review of the model proposed in
\cite{18,19}.  Consider the background manifold $ \overline{V}_{4}
$ isometrically embedded in a pseudo-Riemannian manifold $ V_{m}$
by the map ${ \cal Y}: \overline{V}_{4}\rightarrow  V_{m} $ such
that
\begin{eqnarray}\label{a}
{\cal G} _{AB} {\cal Y}^{A}_{,\mu } {\cal Y}^{B}_{,\nu}=
\bar{g}_{\mu \nu}  , \hspace{.5 cm} {\cal G}_{AB}{\cal
Y}^{A}_{,\mu}{\cal N}^{B}_{a} = 0  ,\hspace{.5 cm}  {\cal
G}_{AB}{\cal N}^{A}_{a}{\cal N}^{B}_{b} =\bar{g}_{ab}= \pm 1,
\end{eqnarray}
where $ {\cal G}_{AB} $  $ ( \bar{g}_{\mu\nu} ) $ is the metric of
the bulk (brane) space  $  V_{m}  (\overline{V}_{4}) $ in
arbitrary coordinates, $ \{ {\cal Y}^{A} \} $   $  (\{ x^{\mu} \})
$  is the  basis of the bulk (brane) and  ${\cal N}^{A}_{a}$ are
$(m-4)$ normal unit vectors orthogonal to the brane. Perturbation
of $\bar{V}_{4}$ in a sufficiently small neighborhood of the brane
along an arbitrary transverse direction $\xi$ is given by
\begin{eqnarray}\label{a1}
{\cal Z}^{A}(x^{\mu},\xi^{a}) = {\cal Y}^{A} + ({\cal
L}_{\xi}{\cal Y})^{A}, \label{eq2}
\end{eqnarray}
where $\cal L$ represents the Lie derivative and $\xi^{a}$ $(a =
1,2,...,m-4)$ is a small parameter along ${\cal N}^{A}_{a}$
parameterizing the extra noncompact dimensions. By taking $\xi$
orthogonal to the brane, we ensure gauge independency \cite{18}
and have perturbations of the embedding along a single orthogonal
extra direction $\bar{{\cal N}}_{a}$, giving the local coordinates
of the perturbed brane as
\begin{eqnarray}\label{a2}
{\cal Z}^{A}_{,\mu}(x^{\nu},\xi^{a}) = {\cal Y}^{A}_{,\mu} +
\xi^{a}\bar{{\cal N}}^{A}_{a,\mu}(x^{\nu}).
\end{eqnarray}
In a similar manner we see that since the vectors $\bar{{\cal
N}}^{A}$ depend only on the local coordinates $x^{\mu}$, they do
not propagate along the extra dimensions. The above  assumptions
lead to the embedding equations of the perturbed geometry
\begin{eqnarray}\label{a4}
g_{\mu \nu }={\cal G}_{AB}{\cal Z}_{\,\,\ ,\mu }^{A}{\cal
Z}_{\,\,\ ,\nu }^{B},\hspace{0.5cm}g_{\mu a}={\cal G}_{AB}{\cal
Z}_{\,\,\ ,\mu
}^{A}{\cal N}_{\,\,\ a}^{B},\hspace{0.5cm}{\cal G}_{AB}{\cal N}_{\,\,\ a}^{A}%
{\cal N}_{\,\,\ b}^{B}={g}_{ab}.
\end{eqnarray}
If we set ${\cal N}_{\,\,\ a}^{A}=\delta _{a}^{A}$, the metric of
the bulk space can be written in the following matrix form
\begin{eqnarray}
{\cal G}_{AB}=\left( \!\!\!%
\begin{array}{cc}
g_{\mu \nu }+A_{\mu c}A_{\,\,\nu }^{c} & A_{\mu a} \\
A_{\nu b} & g_{ab}%
\end{array}%
\!\!\!\right) ,  \label{F}
\end{eqnarray}%
where
\begin{eqnarray}
g_{\mu \nu }=\bar{g}_{\mu \nu }-2\xi ^{a}\bar{K}_{\mu \nu a}+\xi ^{a}\xi ^{b}%
\bar{g}^{\alpha \beta }\bar{K}_{\mu \alpha a}\bar{K}_{\nu \beta
b}, \label{G}
\end{eqnarray}%
is the metric of the perturbed brane, so that
\begin{eqnarray}
\bar{K}_{\mu \nu a}=-{\cal G}_{AB}{\cal Y}_{\,\,\,,\mu }^{A}{\cal
N}_{\,\,\ a;\nu }^{B},  \label{H}
\end{eqnarray}%
represents the extrinsic curvature of the original brane (second
fundamental form). We use the notation $A_{\mu c}=\xi ^{d}A_{\mu
cd}$ where
\begin{equation}
A_{\mu cd}={\cal G}_{AB}{\cal N}_{\,\,\ d;\mu }^{A}{\cal N}_{\,\,\ c}^{B}=%
\bar{A}_{\mu cd},  \label{I}
\end{equation}%
represents the twisting vector fields
(the normal fundamental form). Any fixed $%
\xi ^{a}$ signifies a new perturbed geometry, enabling us to
define an extrinsic curvature similar to the original one by
\begin{eqnarray}
\widetilde{K}_{\mu \nu a}=-{\cal G}_{AB}{\cal Z}_{\,\,\ ,\mu }^{A}{\cal N}%
_{\,\,\ a;\nu }^{B}=\bar{K}_{\mu \nu a}-\xi ^{b}\left( \bar{K}_{\mu \gamma a}%
\bar{K}_{\,\,\ \nu b}^{\gamma }+A_{\mu ca}A_{\,\,\ b\nu
}^{c}\right) . \label{J}
\end{eqnarray}%
Note that definitions (\ref{F}) and (\ref{J}) require
\begin{eqnarray}
\widetilde{K}_{\mu \nu a}=-\frac{1}{2}\frac{\partial {\cal G}_{\mu \nu }}{%
\partial \xi ^{a}}.  \label{M}
\end{eqnarray}%
In geometric language, the presence of gauge fields $A_{\mu a}$
tilts the embedded family of sub-manifolds with respect to normal
vectors ${\cal N} ^{A}$. According to our construction, the
original brane is orthogonal to normal vectors ${\cal N}^{A}.$
However,  equation (\ref{a4})  shows that this is not true for the
deformed geometry. Let us change the embedding coordinates and set
\begin{eqnarray}
{\cal X}_{,\mu }^{A}={\cal Z}_{,\mu }^{A}-g^{ab}{\cal
N}_{a}^{A}A_{b\mu }. \label{mama40}
\end{eqnarray}%
The coordinates ${\cal X}^{A}$ describe a new family of embedded
manifolds whose members are always orthogonal to ${\cal N}^{A}$.
In this coordinates the embedding equations of the perturbed brane
is similar to the original one, described by equation (\ref{a}) so
that ${\cal Y}^{A}$ is replaced by ${\cal X}^{A}$. This new
embedding of the local coordinates are suitable for obtaining
induced Einstein field equations on the brane. The extrinsic
curvature of the perturbed brane then becomes
\begin{eqnarray}
K_{\mu \nu a}=-{\cal G}_{AB}{\cal X}_{,\mu }^{A}{\cal N}_{a;\nu }^{B}=\bar{K}%
_{\mu \nu a}-\xi ^{b}\bar{K}_{\mu \gamma a}\bar{K}_{\,\,\nu b}^{\gamma }=-%
\frac{1}{2}\frac{\partial g_{\mu \nu }}{\partial \xi ^{a}},
\label{mama42}
\end{eqnarray}%
which is the generalized York's relation and shows how the
extrinsic curvature propagates as a result of the propagation of
the metric in the direction of extra dimensions. The components of
the Riemann tensor of the bulk written in the embedding vielbein
$\{{\cal X}^{A}_{, \alpha}, {\cal N}^A_a \}$, lead to the
Gauss-Codazzi equations \cite{27}
\begin{eqnarray}\label{a5}
R_{\alpha \beta \gamma \delta}=2g^{ab}K_{\alpha[ \gamma
a}K_{\delta] \beta b}+{\cal R}_{ABCD}{\cal X} ^{A}_{,\alpha}{\cal
X} ^{B}_{,\beta}{\cal X} ^{C}_{,\gamma} {\cal X}^{D}_{,\delta},
\end{eqnarray}
\begin{eqnarray}\label{a6}
2K_{\alpha [\gamma c; \delta]}=2g^{ab}A_{[\gamma ac}K_{ \delta]
\alpha b}+{\cal R}_{ABCD}{\cal X} ^{A}_{,\alpha} {\cal N}^{B}_{c}
{\cal X} ^{C}_{,\gamma} {\cal X}^{D}_{,\delta},
\end{eqnarray}
where ${\cal R}_{ABCD}$ and $R_{\alpha\beta\gamma\delta}$ are the
Riemann tensors for the bulk and the perturbed brane respectively.
By contracting the Gauss equation (\ref{a5}) on ${\alpha}$ and
${\gamma}$ and defining
\begin{eqnarray}\label{a9}
Q_{\mu\nu}=-g^{ab}\left(K^\gamma_{\mu a}K_{\gamma\nu b}-K_a
K_{\mu\nu b}\right)+\frac{1}{2}\left(K_{\alpha\beta
a}K^{\alpha\beta a}-K_a K^a\right)g_{\mu\nu}, \label{eqq7}
\end{eqnarray}
which is an independently conserved quantity, that is
$Q^{\mu\nu}_{\,\,\,\,\, ;\nu}=0$ \cite{23}. Using the
decomposition of the Riemann tensor into the Weyl curvature, the
Ricci tensor and the scalar curvature, we obtain the $4D$ Einstein
equations as
\begin{eqnarray}\label{a11} G_{\mu\nu}&=&G_{AB} {\cal
X}^{A}_{,\mu}{\cal X}^{B}_{,\nu}+Q_{\mu\nu}-{\cal
E}_{\mu\nu}+\frac{(m-3)}{(m-2)}g^{ab}{\cal R}_{AB}{\cal
N}^{A}_{a}{\cal
N}^{B}_{b}g_{\mu\nu}\nonumber\\
&-&\frac{(m-4)}{(m-2)}{\cal R}_{AB}{\cal X}^{A}_{,\mu}{\cal
X}^{B}_{,\nu}+\frac{(m-4)}{(m-1)(m-2)}{\cal
R}g_{\mu\nu},\label{eqq12}
\end{eqnarray}
where ${\cal E}_{\mu\nu}=g^{ab} C_{ABCD}{\cal N}^{A}_{a}{\cal
X}^{B}_{,\mu}{\cal N}^D_b{\cal X}^C_{,\nu}$ is the electric part
of the Weyl tensor $C_{ABCD}$. Now, let us write the Einstein
field equations in the bulk space
\begin{eqnarray}\label{a13}
G^{(b)}_{AB}+\Lambda^{(b)} {\cal G}_{AB}=\alpha^{*}
S_{AB},\label{eqq14}
\end{eqnarray}
where $\alpha^{*}=\frac{1}{M_{*}^{m-2}}$. In this equation
$\Lambda^{(b)}$ is the cosmological constant of the bulk space
with $S_{AB}$ consisting of two parts
\begin{eqnarray}\label{a14}
S_{AB}=T_{AB}+ \frac{1}{2} {\cal V} {\cal G}_{AB},\label{eqq15}
\end{eqnarray}
where $T_{AB}$ is the energy-momentum tensor of the matter
confined to the brane through the action of the confining
potential $\cal{V}$. We require $\cal{V}$  to satisfy three
general conditions: it should have a deep minimum on the
non-perturbed brane, depends only on extra coordinates $\xi^{a}$
and the gauge group representing the subgroup of the isometry
group of the bulk space should be preserved by it. Although the
explicit form of such a potential is of no consequence in the
present work, for the sake of completeness it would be useful to
mention a result presented in \cite{18} where a confining
potential whose role, as the name suggests, is to trap test
particles on the brane was obtained by invoking the above
properties and assuming that the brane is located at $\xi^{a}=0$,
with the result
\begin{eqnarray}
{\cal V}(\xi)=\frac{1}{2}w^2g_{ab}\xi^{a}\xi^{b},
\end{eqnarray}
where $w$ is a constant much larger than the scale of curvature of
the brane. This potential is clearly of the harmonic oscillator
type, forcing a test particle leaving the brane along the extra
dimension back to its initial position on the brane which
coincides with the location of minimum of the potential. It is
conceivable that other equivalent potentials having different
forms but possessing the same properties and playing the same role
can be envisaged.

Using Einstein equations (\ref{a13}), we obtain
\begin{eqnarray}\label{a15}
{\cal R}_{AB}=-\frac{\alpha^{*}}{(m-2)}{\cal G}_{AB}
S+\frac{2}{(m-2)}\Lambda^{(b)} {\cal G}_{AB}+\alpha^{*} S_{AB},
\end{eqnarray}
and
\begin{eqnarray}\label{a16}
{\cal R}=-\frac{2}{m-2}(\alpha^{*} S-m\Lambda^{(b)}).
\end{eqnarray}
Substituting ${\cal R}_{AB}$ and ${\cal R}$ from the above into
equation (\ref{a11}), we obtain
\begin{eqnarray}\label{a17}
G_{\mu\nu}&=& Q_{ \mu\nu} - {\cal
E}_{\mu\nu}+\frac{(m-3)}{(m-2)}\alpha^{*}g^{ab}S_{ab}g_{\mu\nu}
+\frac{2\alpha^{*}}{(m-2)}S_{\mu\nu} -
\frac{(m-4)(m-3)}{(m-1)(m-2)}\alpha^{*}Sg_{\mu\nu}\nonumber\\
&+&\frac{(m-7)}{(m-1)}\Lambda^{(b)}g_{\mu\nu}.\label{new1}
\end{eqnarray}
As was mentioned in the introduction, localization of matter on
the brane is realized in this model by the action of a confining
potential. This can simply be done by taking
\begin{eqnarray}
\alpha\tau_{\mu\nu} = \frac{2\alpha^{*}}{(m-2)}T_{\mu\nu},
\hspace{.5 cm}T_{\mu a}=0, \hspace{.5 cm}T_{ab}=0,\label{new4}
\end{eqnarray}
where $\alpha$ is the scale of energy on the brane. Now, the
induced Einstein field equations on the original brane can be
written as
\begin{eqnarray}
G_{\mu\nu} = \alpha \tau_{\mu\nu}
-\frac{(m-4)(m-3)}{2(m-1)}\alpha\tau g_{\mu\nu} - \Lambda
g_{\mu\nu} + Q_{\mu\nu} - {\cal E}_{\mu\nu},\label{a8}
\end{eqnarray}
where  $\Lambda= -\frac{(m-7)}{(m-1)} \Lambda^{(b)}$ and
$Q_{\mu\nu}$ is an independently conserved quantity which may be
considered as an energy-momentum tensor of the generalized
Chaplygin gas (gCg). This matter is a candidate for a unified dark
matter-energy scenario which is parameterized by an equation of
state $p=-\frac{B}{\rho^{\beta}}$, where $B$ and $\beta$ are
arbitrary constants. The speed of sound $v_{s}^{2}=-\frac{\partial
p}{\partial\rho}$ in the gCg is defined as $v_{s}^{2}=-\frac{\beta
p}{\rho}$. Ultimately, we have three different types of `matter'
on the right hand side of equation (\ref{a8}), namely, ordinary
confined conserved matter represented by $\tau_{\mu\nu}$, the
matter represented by $Q_{\mu\nu}$ which will be discussed later
and finally, the Weyl matter represented by ${\cal E}_{\mu\nu}$.

The geometrical approach considered here is based on three basic
postulates, namely, the confinement of the standard gauge
interactions to the brane, the existence of quantum gravity in the
bulk and finally, the embedding of the brane world. All other
model dependent properties such as warped metric, mirror
symmetries, radion or extra scalar fields, fine tuning parameters
like the tension of the brane and the choice of a junction
condition are left out as much as possible in our calculations
\cite{24}. In the next section, we discuss the cosmological
implications of our model. As was mentioned above, the results do
not depend on the precise shape of the potential.
\section{Friedmann equations, deceleration parameter and age of the universe}
In what follows we will analyze the influence of the extrinsic
curvature terms on a FRW universe, regarded as a brane embedded in
a $5$-dimensional bulk with constant curvature $({\cal
E}_{\mu\nu}=0)$. The FRW line element is given by
\begin{equation}\label{1}
ds^2=-dt^2+a(t)^2\left[\frac{dr^2}{1-kr^2}+r^2\left(d\theta^2+\sin^2\theta
d\varphi^2\right)\right].
\end{equation}
The confined source is the perfect fluid given in co-moving
coordinates by
\begin{equation}\label{2}
\tau_{\mu\nu}=(\rho+p) u_{\mu}u_{\nu}+pg_{\mu\nu},\hspace{.5
cm}u_{\mu}=-\delta^{0}_{\mu},\hspace{.5 cm}p=(\gamma-1)\rho.
\end{equation}
The constant curvature bulk is characterized by the Riemann tensor
\begin{eqnarray}
{\cal R}_{ABCD}=k_{*}({\cal G}_{AC}{\cal G}_{BD}-{\cal
G}_{AD}{\cal G}_{BC}),\label{1}
\end{eqnarray}
where ${\cal G}_{AB}$ denotes the bulk metric components in
arbitrary coordinates and $k_{*}$ is either zero for the flat
bulk, or proportional to a positive or negative bulk cosmological
constant respectively, corresponding to two possible signatures
$(4,1)$ for the $dS_{5}$ bulk and $(3,2)$ for the $AdS_{5}$ bulk.
We take, in the embedding equations, $g^{55}=\varepsilon=\pm1$.
With this assumption the Gauss-Codazzi equations reduce to
\begin{eqnarray}
R_{\alpha\beta\gamma\delta} = \frac{1}{\varepsilon}
(K_{\alpha\gamma}K_{\beta\delta}-K_{\alpha\delta}K_{\beta\gamma})
+ k_{*}
(g_{\alpha\gamma}g_{\beta\delta}-g_{\alpha\delta}g_{\beta\gamma}),\label{2}
\end{eqnarray}
\begin{eqnarray}\label{3}
K_{\alpha[\beta;\gamma]} = 0.\label{3}
\end{eqnarray}
Also the effective field equations derived in the previous section
with constant curvature bulk can be written as
\begin{eqnarray}\label{4}
G_{\mu\nu} = \alpha \tau_{\mu\nu}-\lambda g_{\mu\nu} + Q_{\mu\nu}.
\end{eqnarray}
Here, $\lambda$ is the effective cosmological constant in four
dimensions with $Q_{\mu\nu}$ being a completely geometrical
quantity given by
\begin{eqnarray}\label{5}
Q_{\mu\nu}=\frac{1}{\varepsilon}\left[\left(KK_{\mu\nu}-
K_{\mu\alpha }K^{\alpha}_{\nu}\right)+\frac{1}{2}
\left(K_{\alpha\beta}K^{\alpha\beta}-K^2\right)g_{\mu\nu}\right],
\end{eqnarray}
where $K=g^{\mu\nu}K_{\mu\nu}$. Using the York's relation
\begin{eqnarray}\label{York}
K_{\mu \nu a}=-\frac{1}{2}\frac{\partial
g_{\mu\nu}}{\partial\xi^{a}},
\end{eqnarray}
we realize that in a diagonal metric, $K_{\mu\nu a}$ is diagonal.
After separating the spatial components, the Codazzi equations
reduce to (here $\alpha,\beta,\gamma,\sigma=1,2,3$)
\begin{eqnarray}\label{6}
K^{\alpha}_{\,\,\,\gamma a,\sigma}+K^{\beta}_{\,\,\,\gamma
a}\Gamma^{\alpha}_{\,\,\,\beta\sigma}= K^{\alpha}_{\,\,\,\sigma
a,\gamma}+K^{\beta}_{\,\,\,\sigma
a}\Gamma^{\alpha}_{\,\,\,\beta\gamma},\label{6}
\end{eqnarray}
\begin{eqnarray}\label{7}
K^{\alpha}_{\,\,\,\gamma
a,0}+\frac{\dot{a}}{a}K^{\alpha}_{\,\,\,\gamma
a}=\frac{\dot{a}}{a}K^{0}_{\,\,\,0a}.\label{7}
\end{eqnarray}
The first equation gives $K^{1}_{\,\,\,1 a,\sigma}=0$  for
$\sigma\neq1$, since $K^{1}_{\,\,\,1 a}$ does not depend on the
spatial coordinates. Repeating the same procedure for
$\alpha,\gamma=i, i=2,3,$ we obtain $K^{2}_{\,\,\,2 a,\sigma}=0$
for $\sigma\neq2$ and $K^{3}_{\,\,\,3 a,\sigma}=0$ for
$\sigma\neq3$. Assuming $K^{1}_{\,\,\,1 a}=K^{2}_{\,\,\,2
a}=K^{3}_{\,\,\,3 a}=b_{a}(t)$, where $b_{a}(t)$ are arbitrary
functions of $t$, the second equation gives
\begin{eqnarray}
K_{00
a}=-\left(\frac{\dot{b}_{a}a}{\dot{a}}+b_{a}\right).\label{8}
\end{eqnarray}
For $\mu,\nu=1,2,3 $ we obtain
\begin{eqnarray}
K_{\mu\nu a}=b_{a}g_{\mu\nu}.\label{9}
\end{eqnarray}
Assuming that the functions $b_{a}$ are equal and denoting
$b_{a}=b$, $h=\frac{\dot{b}}{b}$ and $H=\frac{\dot{a}}{a}$, we
find from equation (\ref{5}) that
\begin{eqnarray}
K_{\alpha\beta}
K^{\alpha\beta}=b^2\left(\frac{h^2}{H^2}+2\frac{h}{H}+4\right)
,\hspace{.5 cm}K=b\left(\frac{h}{H}+4\right),\label{10new}
\end{eqnarray}
\begin{eqnarray}
Q_{\mu\nu}=-\frac{3b^2}{\varepsilon}\left(\frac{2h}{3H}+1\right)g_{\mu\nu},
\hspace{5mm}\mu,\nu=1,2,3,\hspace{.5 cm}
Q_{00}=\frac{3b^2}{\varepsilon}.\label{10}
\end{eqnarray}
As we noted before, $Q_{\mu\nu}$ is an independently conserved
quantity, suggesting the same behavior as that of an
energy-momentum tensor of an uncoupled non-conventional energy
source. With this analogy we take the gCg model as a practical
example and define $Q_{\mu\nu}$ as an exotic fluid and write
\begin{eqnarray}\label{11}
Q_{\mu\nu}\equiv\frac{1}{\alpha}\left[(\rho_{ch}+p_{ch})
u_{\mu}u_{\nu}+p_{ch} g_{\mu\nu}\right] ,\hspace{.5 cm}
p_{ch}=-\frac{B}{\rho_{ch}^{\,\,\,\,\beta}},\label{11}
\end{eqnarray}
where $B$ is a positive constant and $0<\beta\leq1$ \cite{4}.
Comparing $Q_{\mu\nu}, \mu,\nu=1,2,3$ and $Q_{00}$ from equation
(\ref{11}) with the components of $Q_{\mu\nu}$ and $Q_{00}$ given
by equation (\ref{10}), we obtain
\begin{eqnarray}\
p_{ch}=-\frac{3b^2}{\alpha\varepsilon}\left(\frac{2h}{3H}+1\right)
,\hspace{.5 cm}\rho_{ch}=\frac{3b^2}{\alpha\varepsilon}.\label{12}
\end{eqnarray}
Equation (\ref{11}) was chosen in accordance with the weak-energy
condition corresponding to a positive energy density and negative
pressure with $\varepsilon=+1$. Use of the above equations leads
to an equation for $b(t)$
\begin{eqnarray}\label{9}
\left(\frac{2a\dot{b}}{3\dot{a}b}+1\right)=B\left(\frac{3b^2}{\alpha\varepsilon}\right)
^{-(1+\beta)},\label{13}
\end{eqnarray}
for which the solution is
\begin{eqnarray}\label{14}
b(t)=\left(\frac{\alpha\varepsilon}{3}\right)^{\frac{1}{2}}
\left[B+\frac{C}{a^{3(1+\beta)}}\right]^{\frac{1}{2(1+\beta)}},\label{14}
\end{eqnarray}
where $C$ is an integration constant. Using equation (\ref{12})
and this solution, the energy density of the gCg becomes
\begin{eqnarray}\label{15}
\rho_{ch}=\rho_{ch_{0}}\left[B_{s}
+\frac{(1-B_{s})}{a^{3(1+\beta)}}\right]^{\frac{1}{(1+\beta)}},\label{15}
\end{eqnarray}
where $\rho_{ch_{0}}$ is the gCg density at the present time,
$B_{s}=B\rho_{ch_{0}}^{-(1+\beta)}$ is a dimensionless quantity
related to the speed of sound for the gCg today, $v_{s}^{2}=\beta
B \rho_{ch_{0}}^{-(1+\beta)}$ and $C=\rho_{ch_{0}}^{(1+\beta)}-B$.
From equation (\ref{15}), we see that for $B_{s}=0$ the gCg
behaves like matter, $\rho_{ch}\sim a^{-3}$, whereas for $B_{s}=1$
it behaves as a cosmological constant, $\rho_{ch}\sim constant$
and when $0<B_{s}<1$, the model predicts a behavior as that
between a matter phase in the past and a negative dark energy
regime at late times. This particular behavior of the gCg inspired
some authors to propose a unified scheme for the cosmological
``dark sector,'' an interesting idea which has been considered in
many different contexts.

The contracted Bianchi identities in the bulk space,
$G^{AB}_{\,\,\,\,\,\,\,\,\,;A}=0$, using equation (\ref{eqq14}),
imply
\begin{eqnarray}
\left(T^{AB}+\frac{1}{2} {\cal{V}} {\cal {G}}^{AB}\right)_{
;A}=0.\label{24}
\end{eqnarray}
Since the potential $\cal V$ has a minimum on the brane, the above
conservation equation reduces to
\begin{eqnarray}
\tau^{\mu\nu}_{\,\,\,\,\,;\mu}=0.\label{25}
\end{eqnarray}
For a perfect fluid as a confined matter source, the time
evolution of the energy density of the matter is given by
\begin{eqnarray}
\rho=\rho_{0}a^{-3\gamma}.\label{27}
\end{eqnarray}
Taking equation (\ref{4}) and using the geometrical energy density
for $Q_{\mu\nu}$, the modified Friedmann equations on the brane
become
\begin{eqnarray}\label{16}
\left(\frac{\dot{a}}{a}\right)^2+\frac{k}{a^2}=\frac{\lambda}{3}+\frac{\alpha}{3}\rho_{0}a^{-3\gamma}
+\frac{\alpha}{3\varepsilon}\rho_{ch_{0}}\left[B_{s}+\frac{(1-B_{s})}
{a^{3(1+\beta)}}\right]^{\frac{1}{(1+\beta)}},\label{16}
\end{eqnarray}
\begin{eqnarray}\label{17}
\frac{\ddot{a}}{a}=\frac{\lambda}{3}-\frac{\alpha}{6}\rho_{0}a^{-3\gamma}(3\gamma-2)
+\frac{\alpha}{3\varepsilon}\rho_{ch_{0}}
\left[B_{s}+\frac{(1-B_{s})}{a^{3(1+\beta)}}\right]
^{\frac{-\beta}{(1+\beta)}}
\left[B_{s}-\frac{(1-B_{s})}{2a^{3(1+\beta)}}\right].\label{17}
\end{eqnarray}
Now, we rewrite the Friedmann equation (\ref{16}) in terms of the
redshift $z=\frac{a_0}{a}-1$ and of the observational parameter
$\Omega$, as
\begin{eqnarray}\label{18}
H^2=H_{0}^{2}\left\{\Omega_{k}(1+z)^{2}+\Omega_{\lambda}
+\Omega_{m}(1+z)^{3\gamma}+\frac{\Omega_{ch}}{\varepsilon}
\left[B_{s}+(1-B_{s})(1+z)^{3(1+\beta)}\right]^{\frac{1}{1+\beta}}
\right\},\label{18}
\end{eqnarray}
where $H_{0}$ is the present value of the Hubble parameter,
$\Omega_{m}$, $\Omega_{\lambda}$ and $\Omega_{k}$ are
respectively, the confined matter, cosmological constant and
spatial curvature relative density parameters where the latter is
associated with the geometrical dark energy through
$\Omega_{ch}=\frac{\alpha \rho_{ch_{0}}}{3H_{0}^{2}}$. The
deceleration parameter $q$ defined by
$q=-\frac{\ddot{a}a}{\dot{a}^2}$ becomes
\begin{equation}\label{dec}
q=\frac{-\Omega_{\lambda}
+\frac{\Omega_{m}}{2}(1+z)^{3}-\frac{\Omega_{ch}}{\varepsilon}\left[B_{s}+
(1-B_{s})(1+z)^{3(1+\beta)}\right] ^{\frac{-\beta}{(1+\beta)}}
\left[B_{s}-\frac{1}{2}(1-B_{s})(1+z)^{3(1+\beta)}\right]}{\Omega_{k}(1+z)^{2}
+\Omega_{\lambda}
+\Omega_{m}(1+z)^{3}+\frac{\Omega_{ch}}{\varepsilon}\left[B_{s}+(1-B_{s})
(1+z)^{3(1+\beta)}\right]^{\frac{1}{(1+\beta)}}}.
\end{equation}
We consider $\Omega_{\lambda}=0$ and show that, within the context
of the present model, the extrinsic curvature can be used to
account for the accelerated expansion of the universe. Figure 1
shows the behavior of the deceleration parameter as a function of
the redshift for selected values of $\Omega_{m}$, $B_{s}$ and
$\beta$ for signature $\varepsilon=+1$. Note that this behavior is
much dependent on the range of the values that $B_{s}$ can take
and is insensitive to the parameter $\beta$. For having an
accelerating universe with $\Omega_m=0.3$, the value of $B_{s}$
should lie in the range $0.5\leq B_{s}\leq1$. Also the value of
$B_{s}$ determines the acceleration redshift $z_{a}$. For
$B_{s}=1$ and $B_{s}=0.6$ with $\Omega_{m}=0.3$, the accelerated
expansion begins at $z_{a}=0.67$ and $z_{a}=0.07$ respectively.

Had we taken the bulk signature to be $(3,2)$ or $\varepsilon
=-1$, equation (\ref{12}) would have represented a fluid with
negative energy and positive pressure, causing an unexpected
behavior of the expansion of the universe. In order to better
visualize this behavior, the deceleration parameter has been
plotted as a function of the redshift for selected values of
$\Omega_{m}$ and $\Omega_{ch}$ in figure 1.

The age of the universe in FRW models is given by
\begin{equation}\label{0b}
t_{0}^{F}=\frac{1}{H_{0}}\int_{0}^{1}\frac{dx}{\left[\frac{\Omega_{m}}{x}+
(1-\Omega_{m})\right]^\frac{1}{2}},
\end{equation}
where $H_{0}^{-1}=9.8 \times 10^9 \mbox{h}^{-1}$ years and the
dimensionless parameter $h$, according to modern data, is about
0.7. Hence, in the flat matter dominated universe with
$\Omega_{total}=1$ the age of the universe would be only 9.3 Gyr,
whereas the oldest globular clusters yield an age of about 12.5
with an uncertainly of 1.5 Gyr \cite{28}. We find the age of the
universe for our model by direct integration of the Friedmann
equation (\ref{16})
\begin{equation}\label{19}
t_{0}^{B}=\frac{1}{H_{0}}\int_{0}^{1}\frac{dx}{\left\{\Omega_{k}
+\Omega_{m}x^{-3\gamma+2}+
x^2\Omega_{ch}\left[B_{s}+\frac{(1-B_{s})}{x^{3(1+\beta)}}\right]
^{\frac{1}{1+\beta}}\right\}^\frac{1}{2}},\label{19}
\end{equation}
where
\begin{eqnarray}\label{20}
\Omega_{k}=-\frac{k}{H_{0}^{2}},\hspace{.5
cm}\Omega_{m}=\frac{\alpha\rho_{0m}}{3H_{0}^{2}},\hspace{.5 cm}
\Omega_{ch}=\frac{\alpha\rho_{ch_{0}}}{3H_{0}^{2}},\label{20}
\end{eqnarray}
For a flat, matter dominated universe with $\Omega_{m}=0.3$ and
$B_{s}=1$ this leads to a prediction for the age of the universe
of about 13.2 Gyr. It shows that the age of the universe in our
model is longer than the FRW model. In figure 2 we have plotted
the dimensionless age parameter $H_{0}t_{0}$ as a function of
$\Omega_{m}$ for some selected values of $B_{s}$ and $\beta$. This
behavior shows that for a fixed value of $\Omega_{m}$ the
predicted age of the universe is longer for larger values of
$B_{s}$. Note also that the age parameter $H_{0}t_{0}$ is an
almost insensitive function of parameter $\beta$ but that it
depends strongly on variations of $B_{s}$. This means that age
considerations will be much more efficient for imposing
constraints on the speed of sound, $B_{s}$, than the values of the
parameter $\beta$ \cite{29}.

Finally, it would be appropriate to compare the predicted age of
the universe in our model to the Randall-Sundrum brane model where
the effects of the brane parameters and dark energy on the age of
the universe have been studied. It has been shown that the effect
of the quadratic term $\rho^2$, resulting from the imposition of
the Israel junction conditions on the energy density term is to
lower significantly the age of the universe. This term effectively
contributes as a positive pressure, making brane models less
accelerating. This problem can be avoided if we accept the
existence of dark energy with $p=-\frac{4}{3}\rho$ (phantom
matter) on the brane since it has a very strong influence on
increasing the age of the universe \cite{30}.
\begin{figure}
\centerline{\begin{tabular}{ccc}
\epsfig{figure=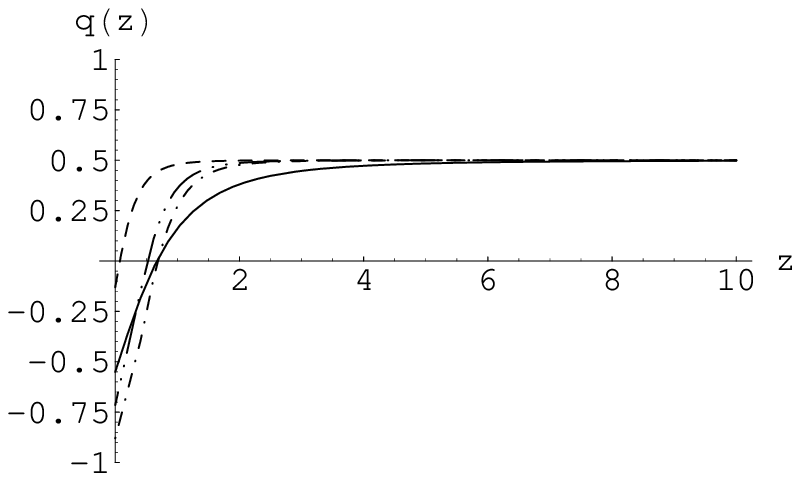,width=7cm}\hspace{1cm}
\epsfig{figure=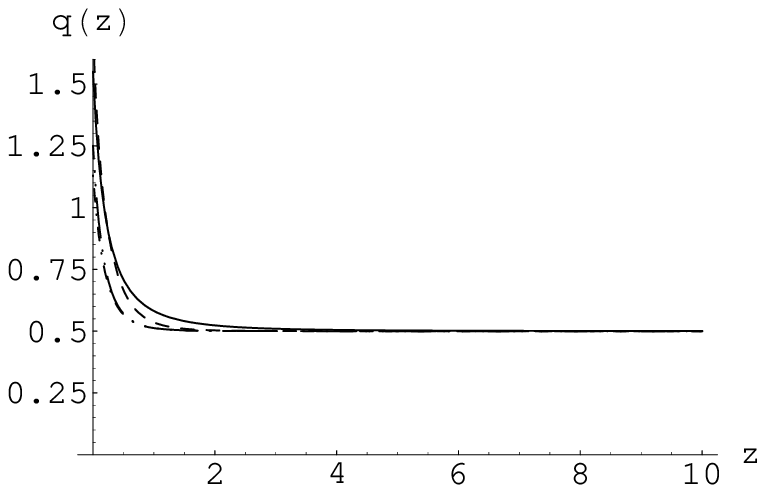,width=7cm}
\end{tabular} } \caption{\footnotesize Left, deceleration parameter as a function of the redshift for $\Omega_{m}=0.3$ and
$B_{s}=1$ (solid line), $\Omega_{m}=0.3$ and $B_{s}=0.6$ (dashed
line), $\Omega_{m}=0$ and $B_{s}=0.92$ (dot-dashed line),
$\Omega_{m}=0.1$ and $B_{s}=0.9$ (dot-dot-dashed line) for
signature $\varepsilon=+1$ with $\beta=1$ and right, the same
parameter for signature $\varepsilon=-1$.}\label{fig1}
\end{figure}
\begin{figure}
\centerline{\begin{tabular}{ccc}
\epsfig{figure=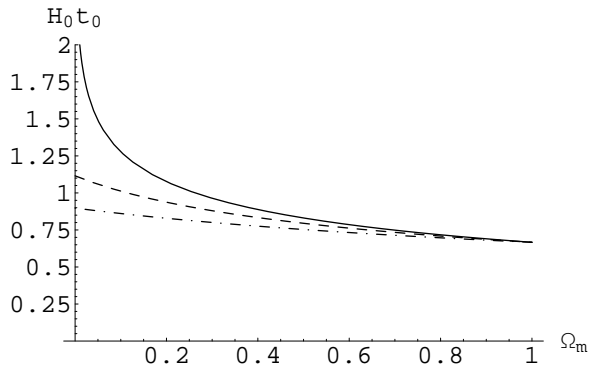,width=8cm}
\end{tabular} } \caption{\footnotesize $H_{0}t_{0}$ as a function
of $\Omega_{m}$ for $B_{s}=1$ (solid line), $B_{s}=0.95$ (dashed line), $B_{s}=0.8$
(dot-dashed line) and $\beta=1$.}\label{fig2}
\end{figure}
\section{Conclusions}
In this paper, we have studied a brane world model in which the
matter is confined to the brane through the action of a confining
potential, rendering the use of any junction condition redundant.
We have shown that in a FRW universe embedded in a constant
curvature $dS_{5}$ bulk the accelerated expansion of the universe
can be explained in a purely geometrical fashion based on the
extrinsic curvature. We have established a correspondence with the
phenomenological gCg dark energy model and have extended the
predictions of geometrical matter in the more general case where
the relation between $p$ and $\rho$ is not linear. Finally, we
have found that the age of the universe in our model is longer
than that predicted by the FRW models and hence more in line with
the present observational data.

\end{document}